\documentclass{moriond}
\usepackage{xspace}
\usepackage{amsmath}
\usepackage{lineno}
\def\Journal#1#2#3#4{{#1} {\bf #2}, #3 (#4)}

\def\PRD{{\em Phys. Rev.} D}

\def\be{\begin{equation}}
\def\ee{\end{equation}}
\def\bea{\begin{eqnarray}}
\def\eea{\end{eqnarray}}

\def\qsq{\ensuremath{q^{2}}\xspace}
\def\Bz{\ensuremath{B^{0}}\xspace}
\def\Kstarz{\ensuremath{K^{\ast{}0}}\xspace}
\def\mumu{\ensuremath{\mu^{+}\mu^{-}}\xspace}

\def\ctl {\ensuremath{\cos\theta_{\ell}}\xspace}
\def\cph {\ensuremath{\cos\phi}\xspace}
\def\csqtk {\ensuremath{\cos^{2}\theta_{K}}\xspace}

\def \thk {\ensuremath{\theta_{K}}\xspace}
\def \thl {\ensuremath{\theta_{\ell}}\xspace}

\def \stl {\ensuremath{\sin\theta_{\ell}}\xspace}
\def \sph {\ensuremath{\sin\phi}\xspace}
\def \ssqtk {\ensuremath{\sin^{2}\theta_{K}}\xspace}
\def \ssqtl {\ensuremath{\sin^{2}\theta_{\ell}}\xspace}

\def \jpsi {\ensuremath{J/\psi}\xspace}

\newcommand{\Photo}{}

\begin{document}
\vspace*{4cm}
\title{$b\to{}s\ell\ell$ decays at LHCb}

\author{ M. Smith, on behalf of the LHCb collaboration }

\address{Blackett Laboratory, Imperial College London\\
Prince Consort Road, London SW7 2BW, England}

\maketitle\abstracts{
Flavour changing neutral currents are suppressed in the Standard Model, making them a prime avenue to search for new physics. Although several measurements of the decay \mbox{$B^{0}\to{}K^{\ast{}0}\mu^{+}\mu^{-}$} have shown deviations from the Standard Model expectations, long-distance contributions may be imitating new physics. Here, the first amplitude analysis of \mbox{$B^{0}\to{}K^{\ast{}0}\mu^{+}\mu^{-}$} is carried out to try and discern such effects.}

\section{Introduction}

\begin{figure}[h]
\centering
\includegraphics[width = 0.48\textwidth]{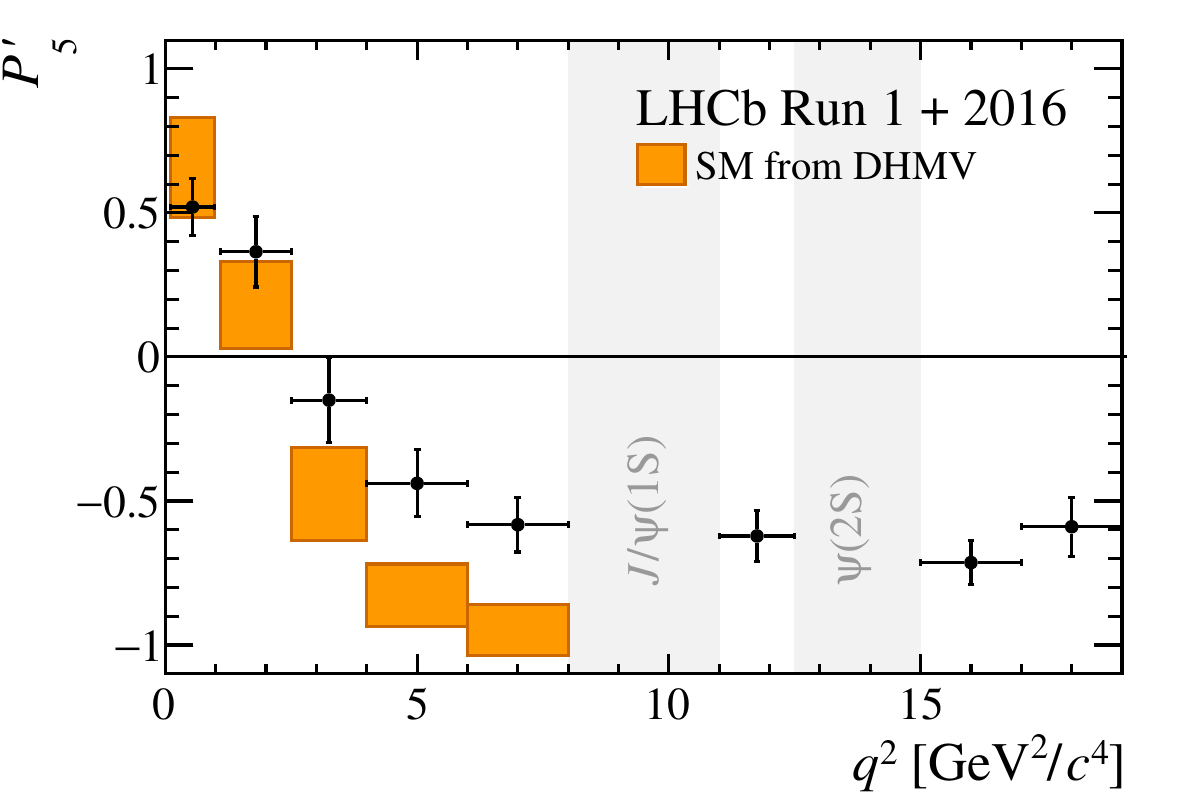}
\caption{The angular observable $P_{5}^{\prime}$ for $\Bz\to\Kstarz\mumu$ as measured by LHCb with the Run~1 and 2016 data sets~\protect\cite{LHCbbinned}. The results are compared with theory predictions~\protect\cite{khodjamirian}~\protect\cite{dmhv}.}
\label{sec:intro:fig:p5p}
\end{figure}

In the Standard Model (SM) flavour changing neutral currents (FCNC) are forbidden at tree-level, leading to small expected amplitudes. Such decays are therefore an excellent place to look for new physics. Several measurements of decays of the form $b\to{}s\ell\ell$ have been made with a sm\"{o}rgasbord of initial and final state hadrons. These analyses include the differential branching fractions of semi leptonic decays~\cite{phimumu,lbmumu,bfs}, tests of lepton universality with final states with electrons or muons~\cite{RKone,RKtwo}, or measurements of angular coefficients of the distribution of the decay products~\cite{LHCbbinned,binnedkstplus}. Whilst recent results for lepton universality show no deviation from the SM, differential branching fractions and angular observables have shown persistent deviations from the SM expectations. An example is the angular observable $P_{5}^{\prime}$ from $\Bz\to\Kstarz\mumu$, shown in Fig.~\ref{sec:intro:fig:p5p}.

The various results may be combined in a fit for the Wilson coefficients of the effective theory that describes such transitions~\cite{eft}. The relevant Wilson coefficients for these decays are $C_{9}^{(\prime)}$ and $C_{10}^{(\prime)}$. The results of the fits show a significant deviation of $C_{9}$ from the Standard Model~\cite{fitone,fittwo,fitthree,fitfour,Gubernari_2022}. However, this is not unambiguous evidence of new physics. The contribution of intermediate $c\bar{c}$ loops, referred to as a `long-distance' or `non-local' effect, gives a contribution to the amplitudes that is identical to the $C_{9}$ Wilson coefficient. These $c\bar{c}$ loops are difficult to calculate and so a source of significant uncertainty in the theory predictions. In this analysis the experimental data has been used to try and estimate the size of these long-distance effects in $B^{0}\to{}K^{\ast0}\mu^{+}\mu^{-}$~\cite{ampone,amptwo}.

\section{Analysis}

The decay $\Bz\to\Kstarz\mumu$ is completely described by three decay angles, $\Omega = (\cos\theta_{\ell},\cos\theta_{K},\phi)$, and the invariant mass squared of the two muons, \qsq. The differential decay rate is expressed as the sum of angular terms
\begin{align*}
    \frac{{\rm d}^{4}\Gamma_{P}}{{\rm d}\qsq{\rm d}\Omega} = \frac{9}{32\pi}\bigl[ &J_{1s}\ssqtk + J_{1c}\csqtk 
    + (J_{2s}\ssqtk + J_{2c}\csqtk)\cos2\thl\\
    + &J_{3}\ssqtk\ssqtl\cos{}2\phi  + J_{4}\sin{}2\thk\sin2\thl\cph + J_{5}\sin2\thk\stl\cph\\
    + &J_{6s}\ssqtk\ctl + J_{7}\sin2\thk\stl\sph + J_{8}\sin2\thk\sin2\thl\sph\\
    + &J_{9}\ssqtk\ssqtl\sin2\phi \bigr],
\end{align*}
where the $J_{i}$ coefficients are dependent on \qsq. The angular coefficients are combinations of the \qsq -dependent amplitudes $\mathcal{A}_{\lambda}^{L,R}$, where $\lambda=\parallel,\perp,0$ labels the polarisation state of the \Kstarz. These amplitudes themselves are functions of the Wilson coefficients that describe the short-distance physics of the $b\to{}s\mu\mu$ transition. They also depend on the local form factors, $\mathcal{F}_{\lambda}$ that describe the hadronic physics of $B\to\Kstarz$ and non-local form factors, $\mathcal{H}_{\lambda}$, that parametrise the effect of the $c\bar{c}$ loops. For example
\begin{align*}
\mathcal{A}_{\parallel} \sim \biggl\{ \bigl[ \left( C_{9} + C_{9}^{\prime} \right) - \left( C_{10} + C_{10}^{\prime} \right) \bigr] \mathcal{F}_{\parallel}(\qsq) + \frac{2m_{b}M_{B}}{\qsq} \bigl[ 	\left(C_{7} + C_{7}^{\prime}\right)\mathcal{F}_{\parallel}^{T}(\qsq) - 16\pi^{2}\frac{M_{B}}{m_{b}}\mathcal{H}_{\parallel}(\qsq)\bigr] \biggr\},
\end{align*}
where $m_{b}$ and $M_{B}$ are the masses of the $b$-quark and $\Bz$ meson, respectively. This analysis fits data to directly extract the Wilson coefficients $C_{9,10}^{(\prime)}$ and the local and non-local form factors. The $C_{7}^{(\prime)}$ Wilson coefficients describe photon transitions and are fixed to their SM values~\cite{Bobeth_2000,Gorbahn_2005}.

The local form factors are constrained to determinations from lattice QCD~\cite{Horgan_201530} and light cone sum rules (LCSR)~\cite{Gubernari_2019}. The non-local form factors are parametrised by~\cite{Gubernari_2022}
\begin{align*}
\mathcal{H}_{\lambda}(z) = \frac{1 - z z_{\jpsi}}{z - z_{\jpsi}}\frac{1 - z z_{\psi(2S)}}{z - z_{\psi(2S)}}\hat{\mathcal{H}}_{\lambda}(z), && \hat{\mathcal{H}}_{\lambda}(z) = \phi^{-1}_{\lambda}(z)\sum_{k}a_{\lambda,k}z^{k} .
\end{align*} 
The variable $z$ is a mapping of \qsq defined by
\begin{align*}
\qsq \mapsto z(\qsq) \equiv \frac{\sqrt{t_{+} - \qsq} - \sqrt{t_{+} - t_{0}}}{\sqrt{t_{+} - \qsq} + \sqrt{t_{+} - t_{0}}},
\end{align*}
where $t_{+} = 4M_{D}$, with $M_{D}$ being the $D^{0}$ mass, and $t_{0}$ chosen such that $z(\qsq = t_{0})=0$. The non-local form factors are thus an expansion in powers of the variable $z$. Measurements of $\Bz\to\Kstarz\jpsi$ and $\Bz\to\Kstarz\psi(2S)$~\cite{PhysRevD.90.112009,PhysRevD.76.031102,PhysRevD.88.074026,PhysRevD.88.052002,Aaij_2012} are used to constrain the non-local functions at the poles. They have also been calculated with LCSR in the region $\qsq<0\,{\rm GeV^{2}}$~\cite{Gubernari_2021}, which may be used to add further constraints in the fit.

The data analysed are those collected by LHCb in Run~1 at $\sqrt{s} = 7,8\,{\rm TeV}$ and 2016 at $\sqrt{s} = 13\,{\rm TeV}$, corresponding to an integrated luminosity of $4.7\,{\rm fb}^{-1}$. This is the same data set as for the previous LHCb publication for $\Bz\to\Kstarz\mumu$~\cite{LHCbbinned}. At very low \qsq the data are dominated by decays with intermediate light resonances. At high \qsq $c\bar{c}$ resonances dominate. Therefore a region in \qsq is selected below the \jpsi resonance, $1.1<\qsq<8.0\,{\rm GeV}^{2}$ and between the \jpsi and $\psi(2S)$ resonances $11.0<\qsq<12.5\,{\rm GeV}^{2}$. In this analysis the $\Kstarz$ is reconstructed from $\Kstarz\to K^{+}\pi^{-}$.

The invariant mass of all the final state particles, $m(K^{+}\pi^{-}\mumu)$, is included in the fit to separate signal from backgrounds consisting of random combinations of tracks. The invariant mass of the $K^{-}\pi^{+}$ pair is also in the fit to separate out the contribution of the S-wave $\Bz\to{}K^{+}\pi^{-}\mumu$ decay. In total there are therefore six variables in the fit: $m(K^{+}\pi^{-}\mumu)$, $m(K^{+}\pi^{-})$, $\cos\theta_{\ell}$, $\cos\theta_{K}$, $\phi$ and $\qsq$. The warping of the angular distributions due to the detector acceptance, track reconstruction and selection are modelled with large simulation samples and its effect is included in the fit following the procedure in the previous LHCb analysis~\cite{LHCbbinned}. The fit is carried out twice; once with the LCSR constraints for the non-local form factors at $\qsq<0\,{\rm GeV^2}$ and once without.

\section{Results}

\begin{figure}[]
\centering
\includegraphics[width = 0.32\textwidth]{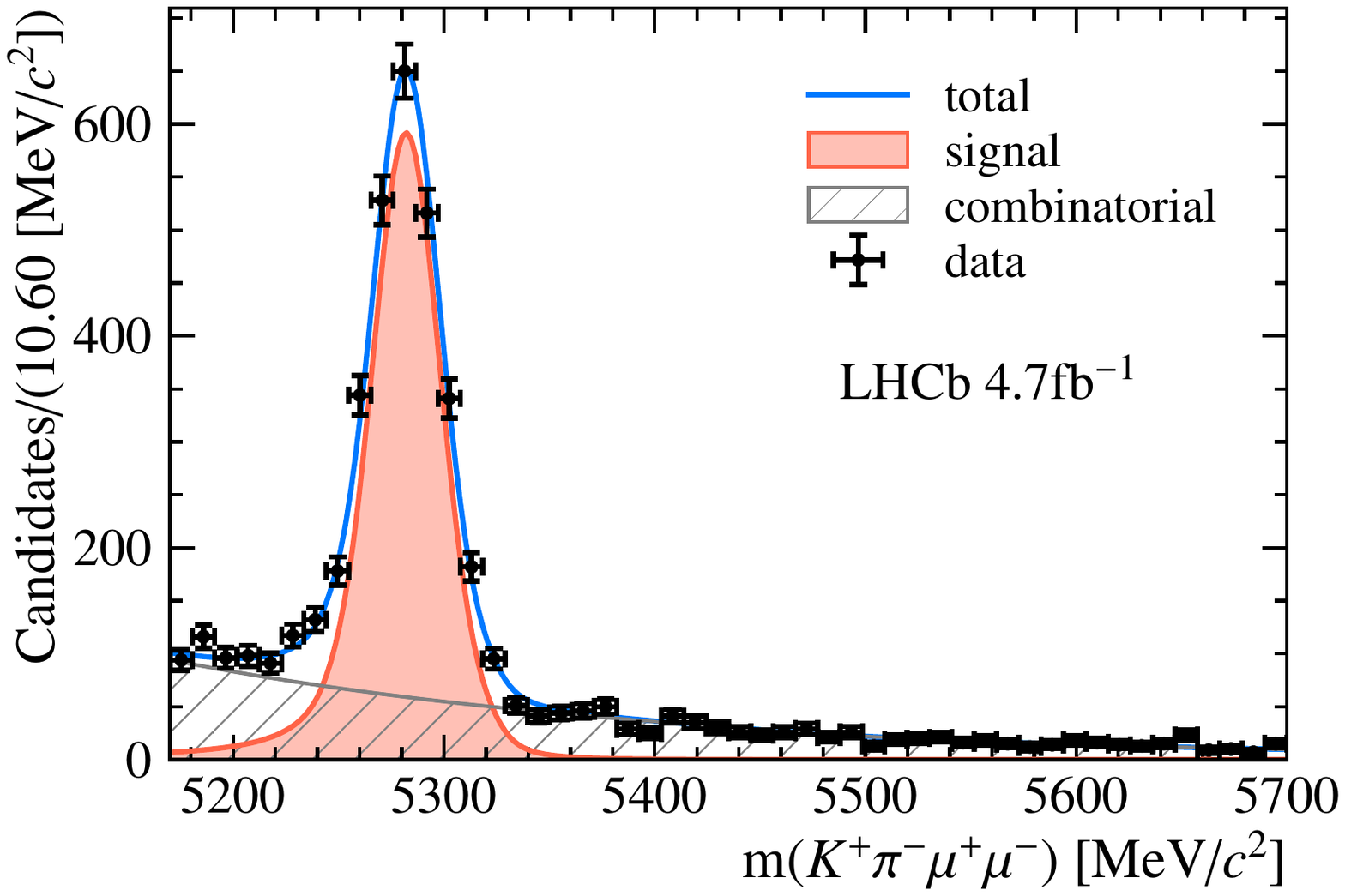}
\includegraphics[width = 0.32\textwidth]{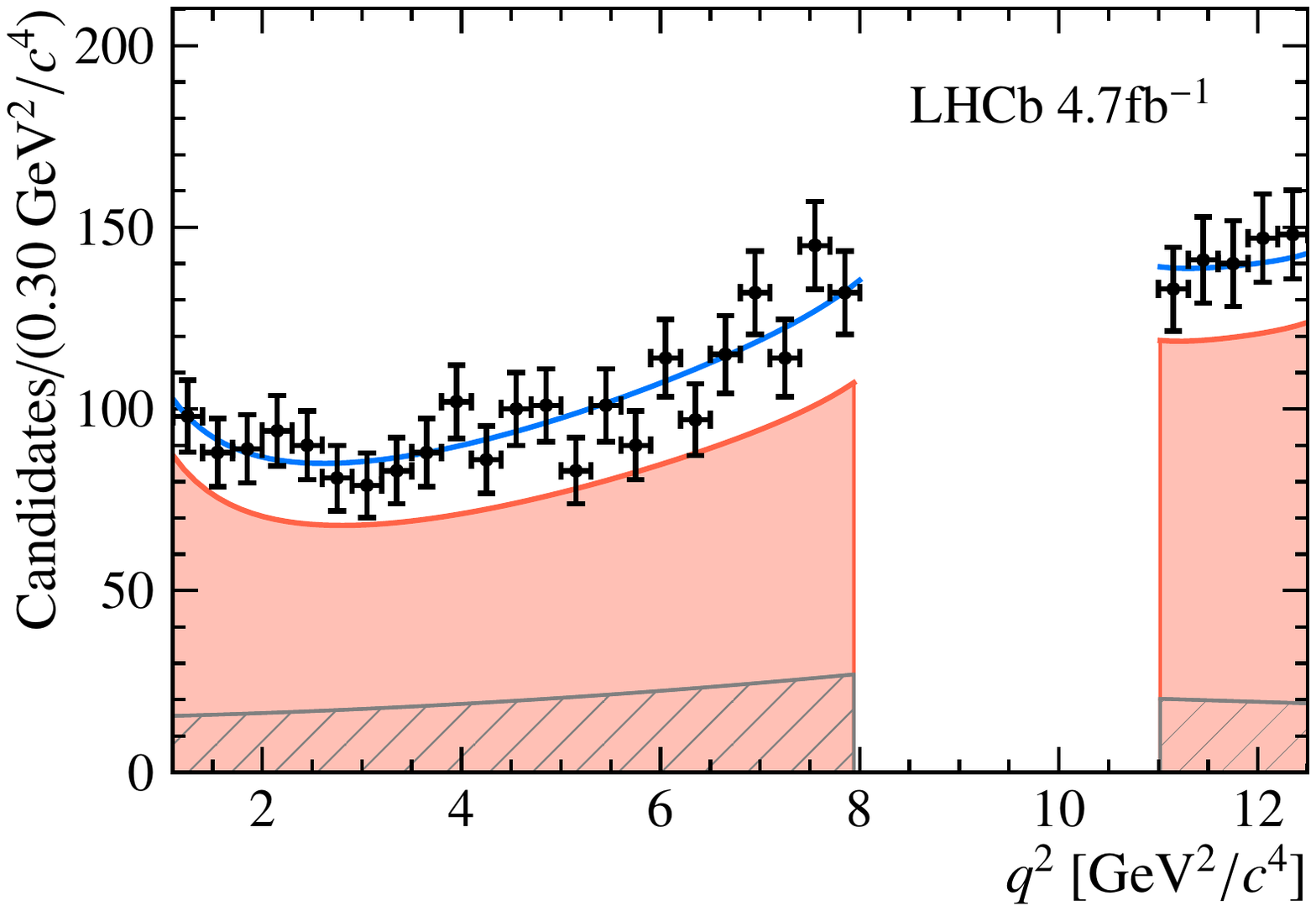}
\includegraphics[width = 0.32\textwidth]{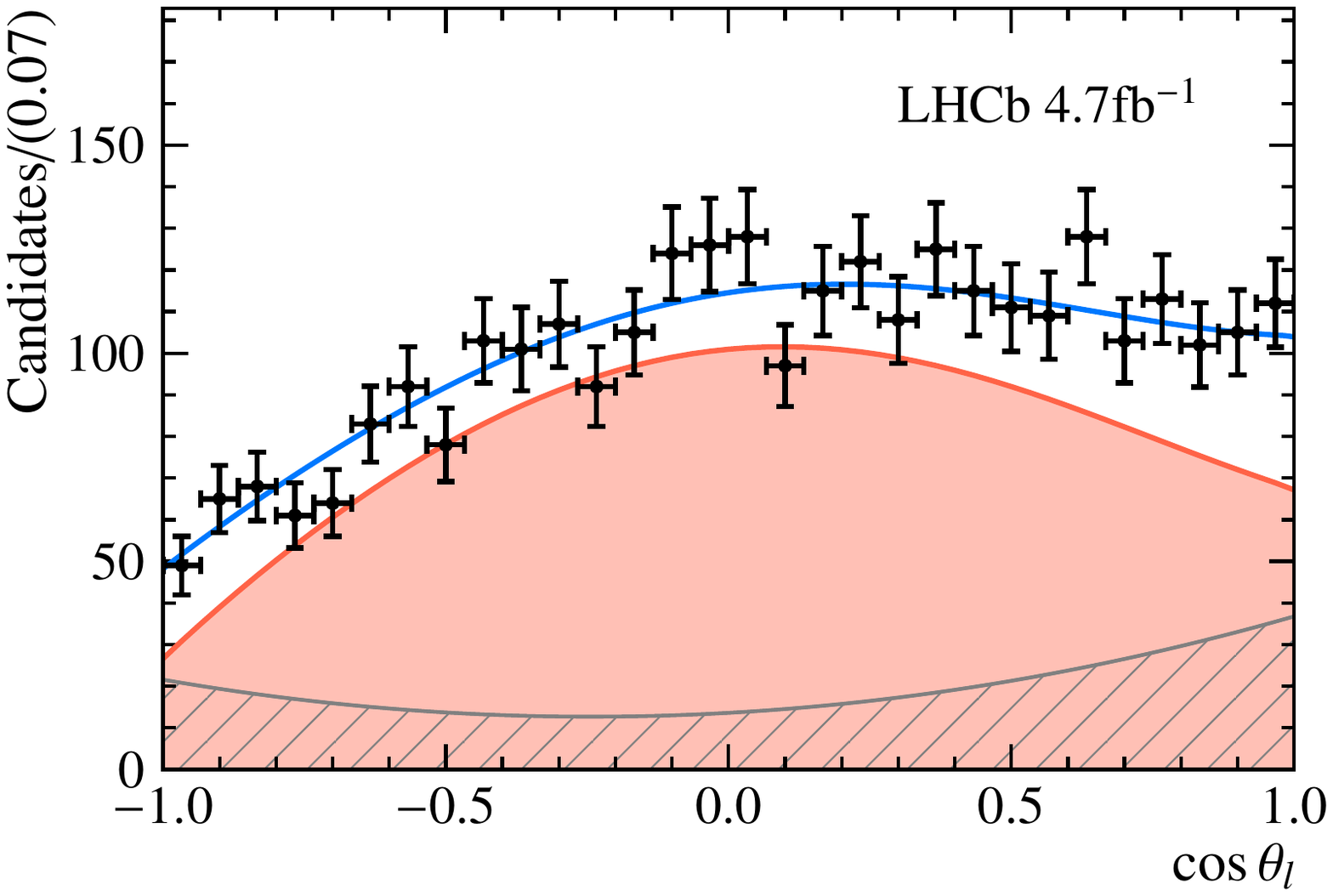}\\
\includegraphics[width = 0.32\textwidth]{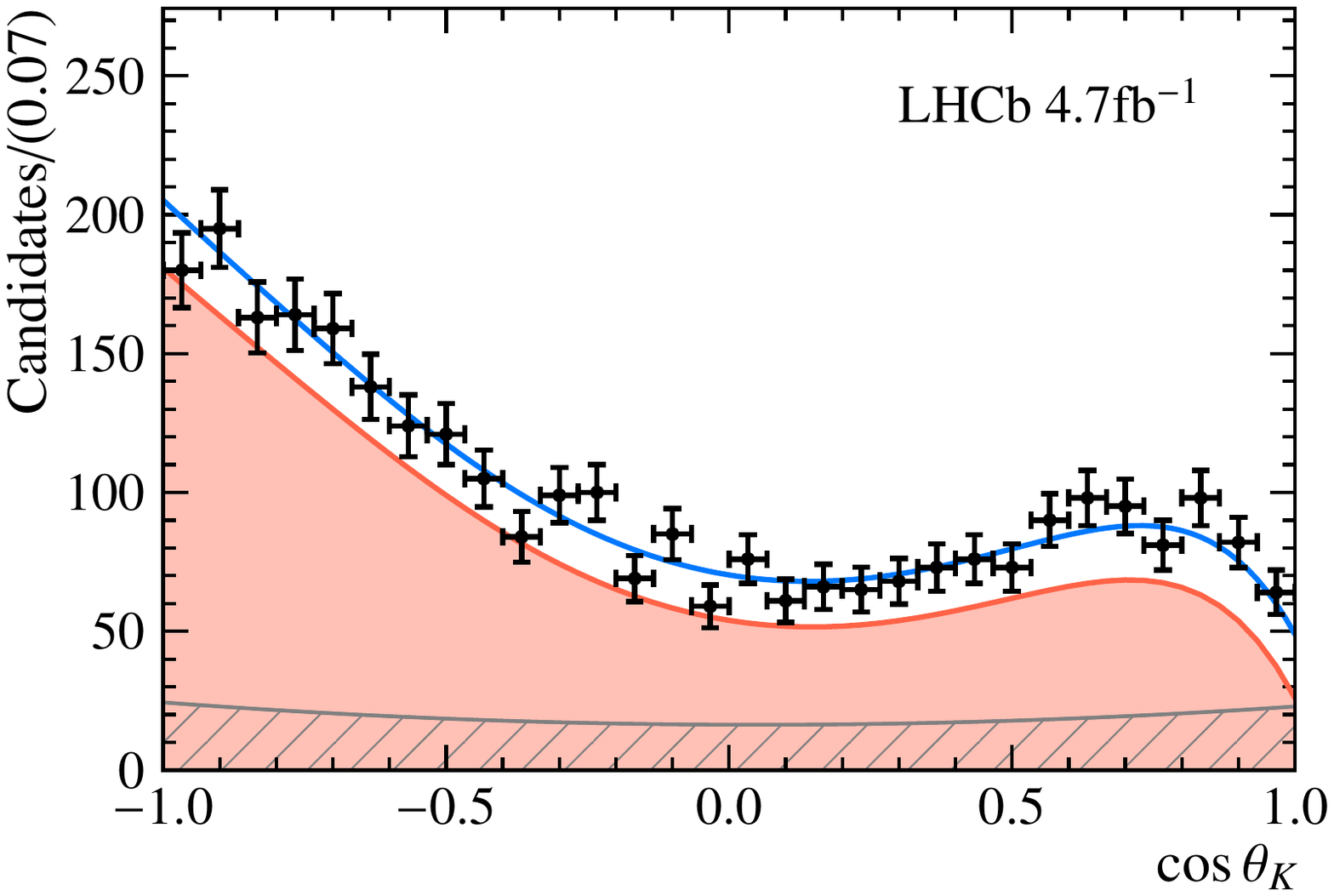}
\includegraphics[width = 0.32\textwidth]{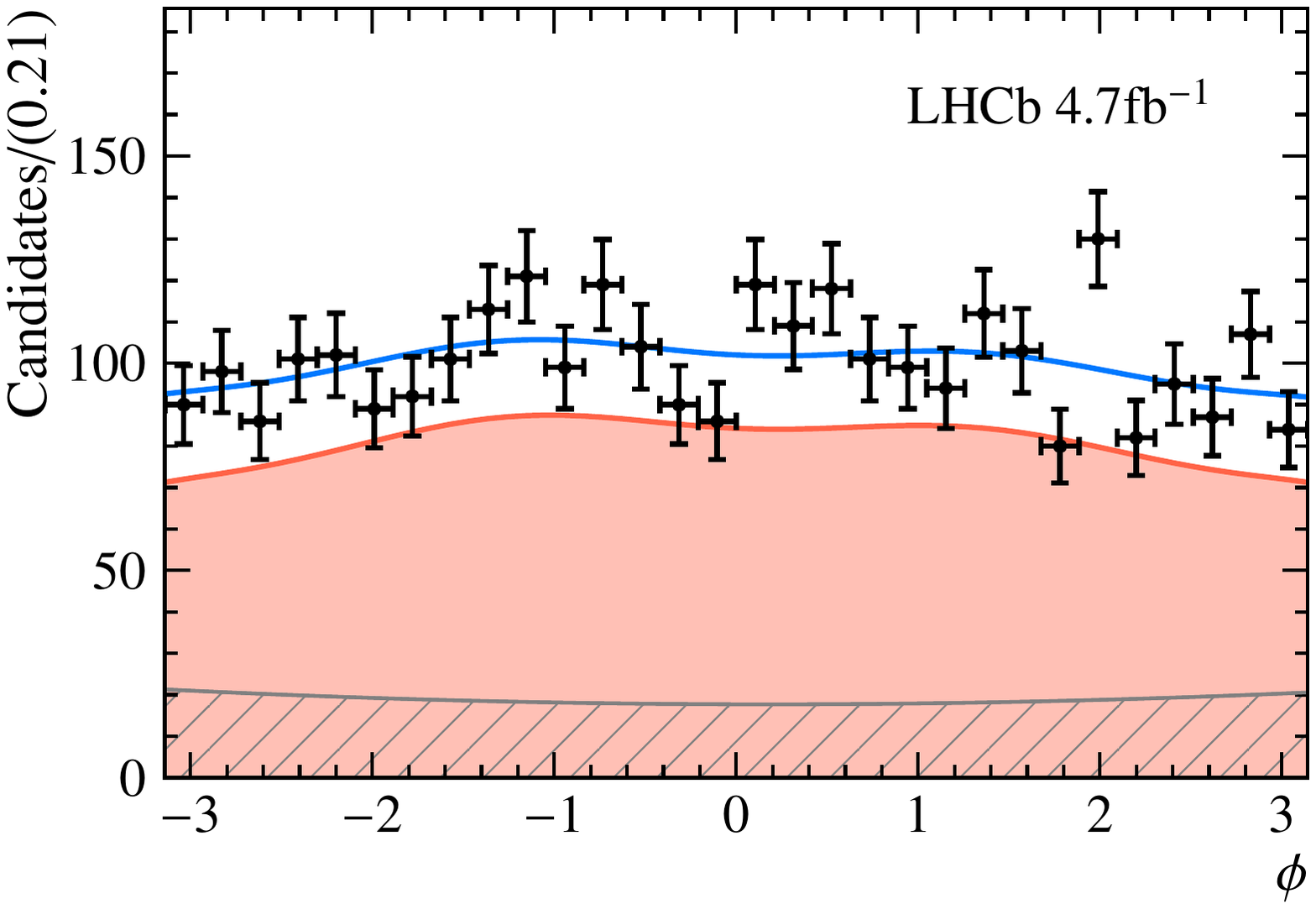}
\includegraphics[width = 0.32\textwidth]{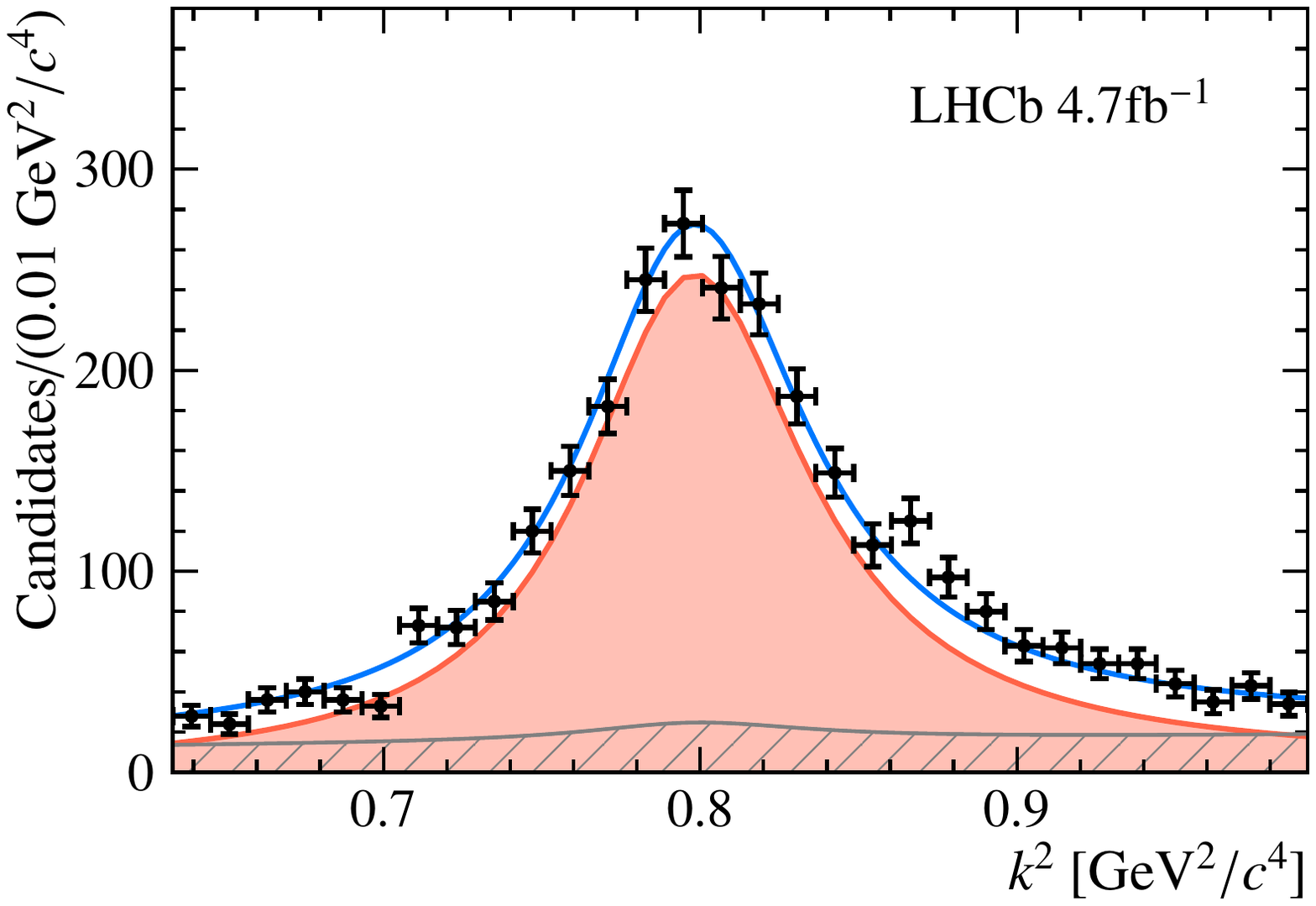}
\caption{Projections of the data in the fit variables with the fit result overlaid.}
\label{sec:results:fig:projections}
\end{figure}

The projections of the data with the fit result overlaid are shown in Fig.~\ref{sec:results:fig:projections}. The fit projections show no discernible change when fitting with or without the negative \qsq constraints.

\begin{figure}[]
\centering
\includegraphics[width = 0.32\textwidth]{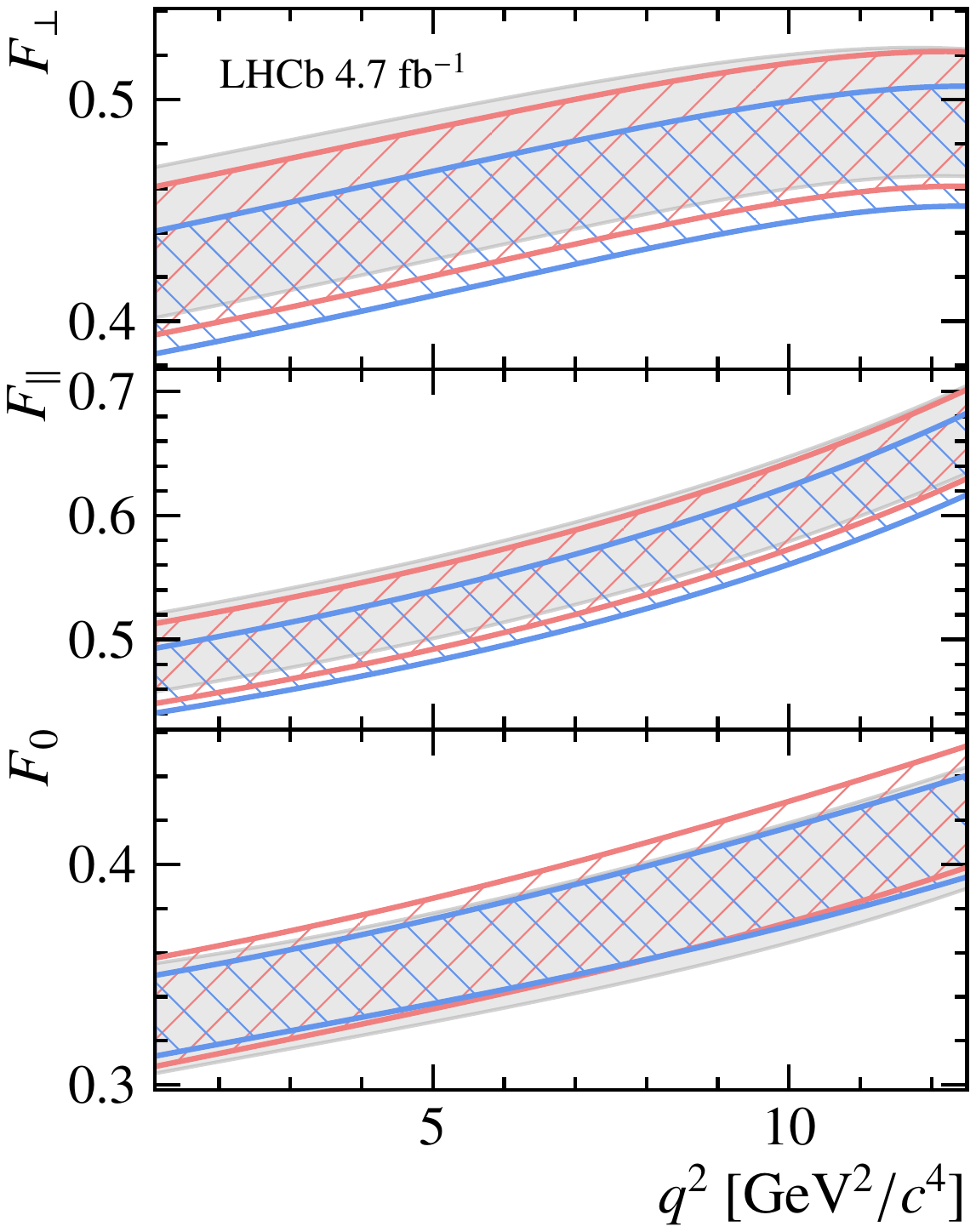}
\includegraphics[width = 0.32\textwidth]{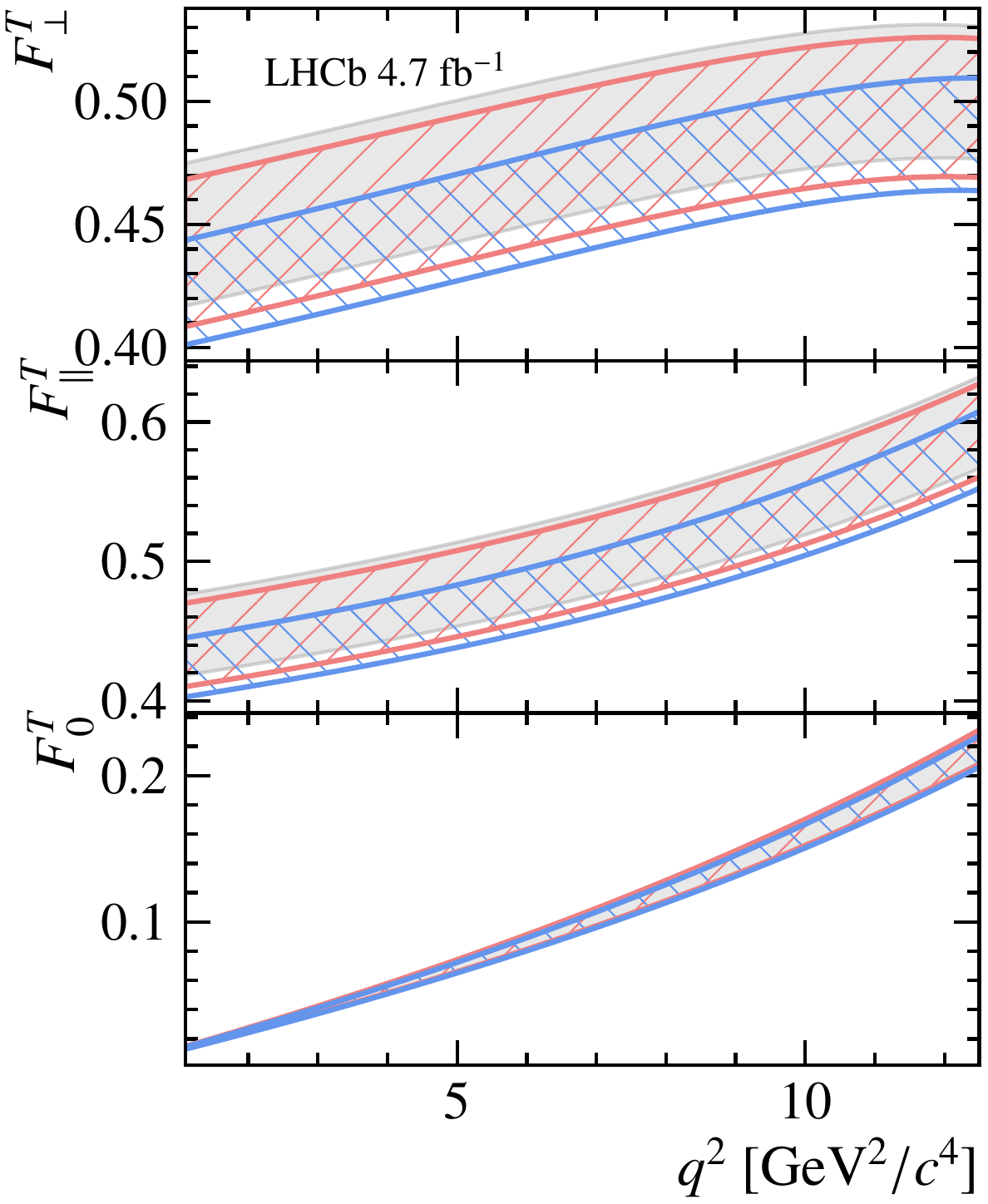}
\includegraphics[width = 0.32\textwidth]{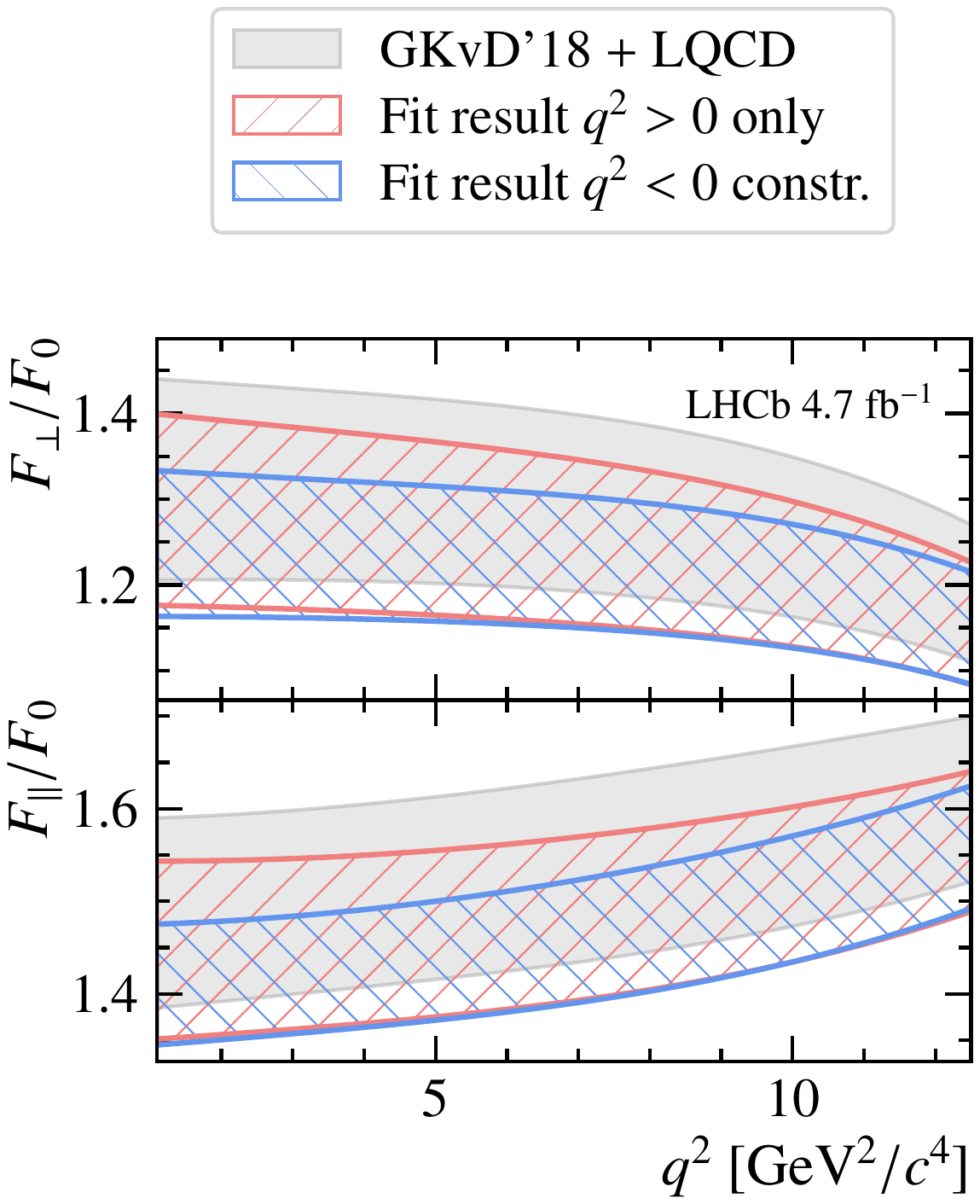}
\caption{The fitted local form factors. The blue curves are the fit with the negative \qsq constraints, the red curve is the fit without.}
\label{sec:results:fig:localffs}
\end{figure}

The fitted local form-factors are shown in Fig.~\ref{sec:results:fig:localffs}. The two fits are consistent with each other and broadly in agreement with the theory prediction. However it is noticeable in the ratios of form factors that the data prefers to pull down the values, compared to the predictions.

\begin{figure}[]
\centering
\includegraphics[width = 0.32\textwidth]{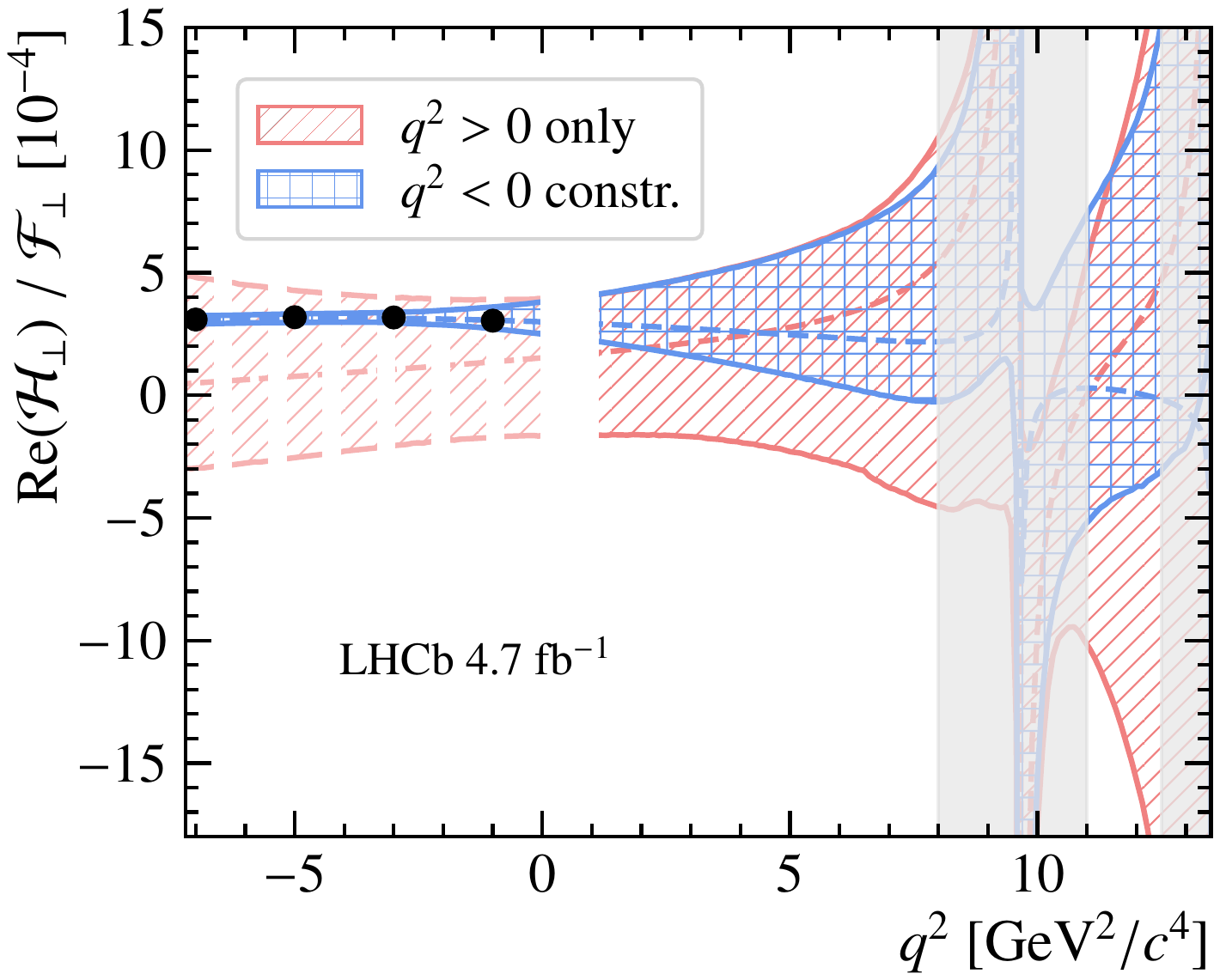}
\includegraphics[width = 0.32\textwidth]{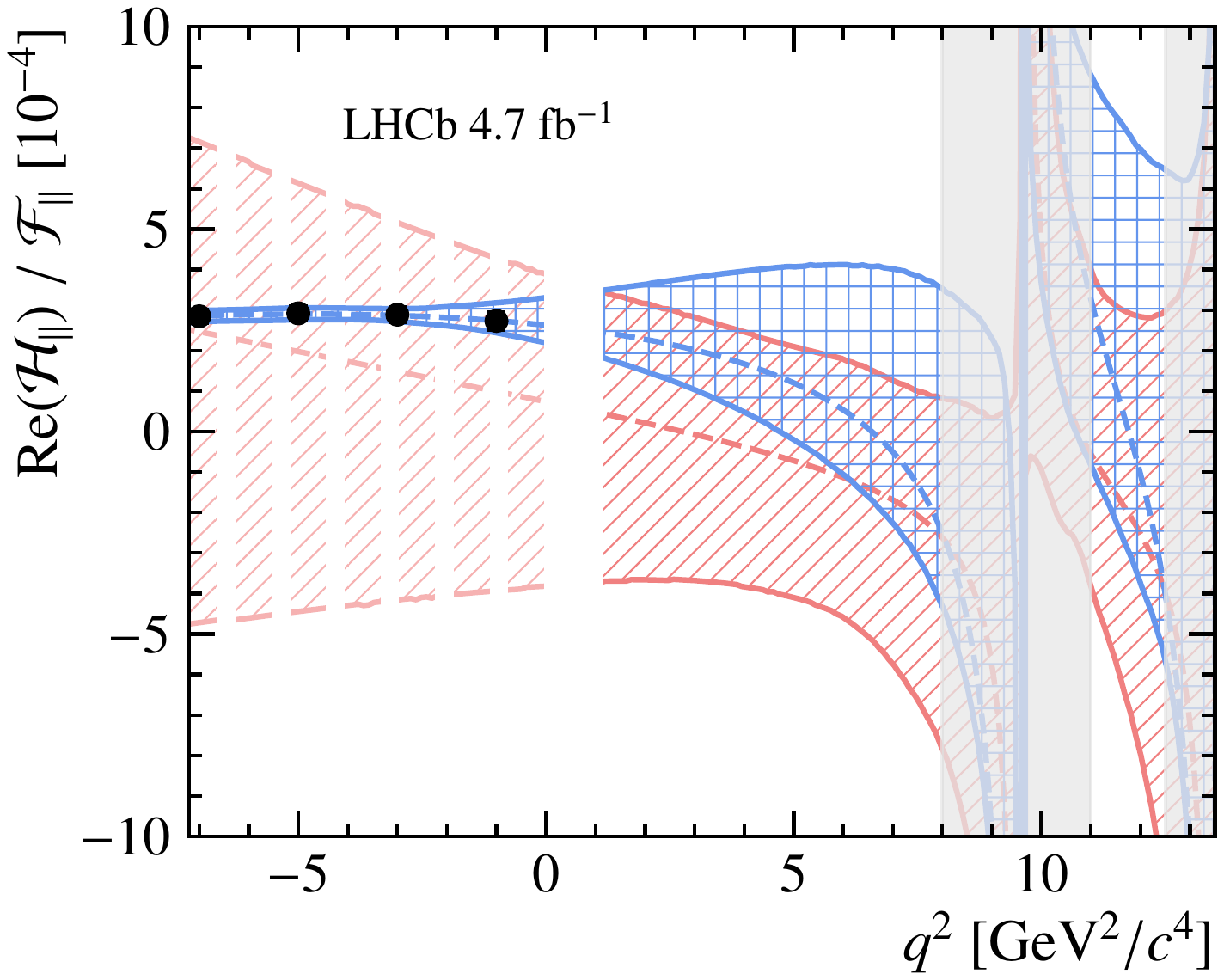}
\includegraphics[width = 0.32\textwidth]{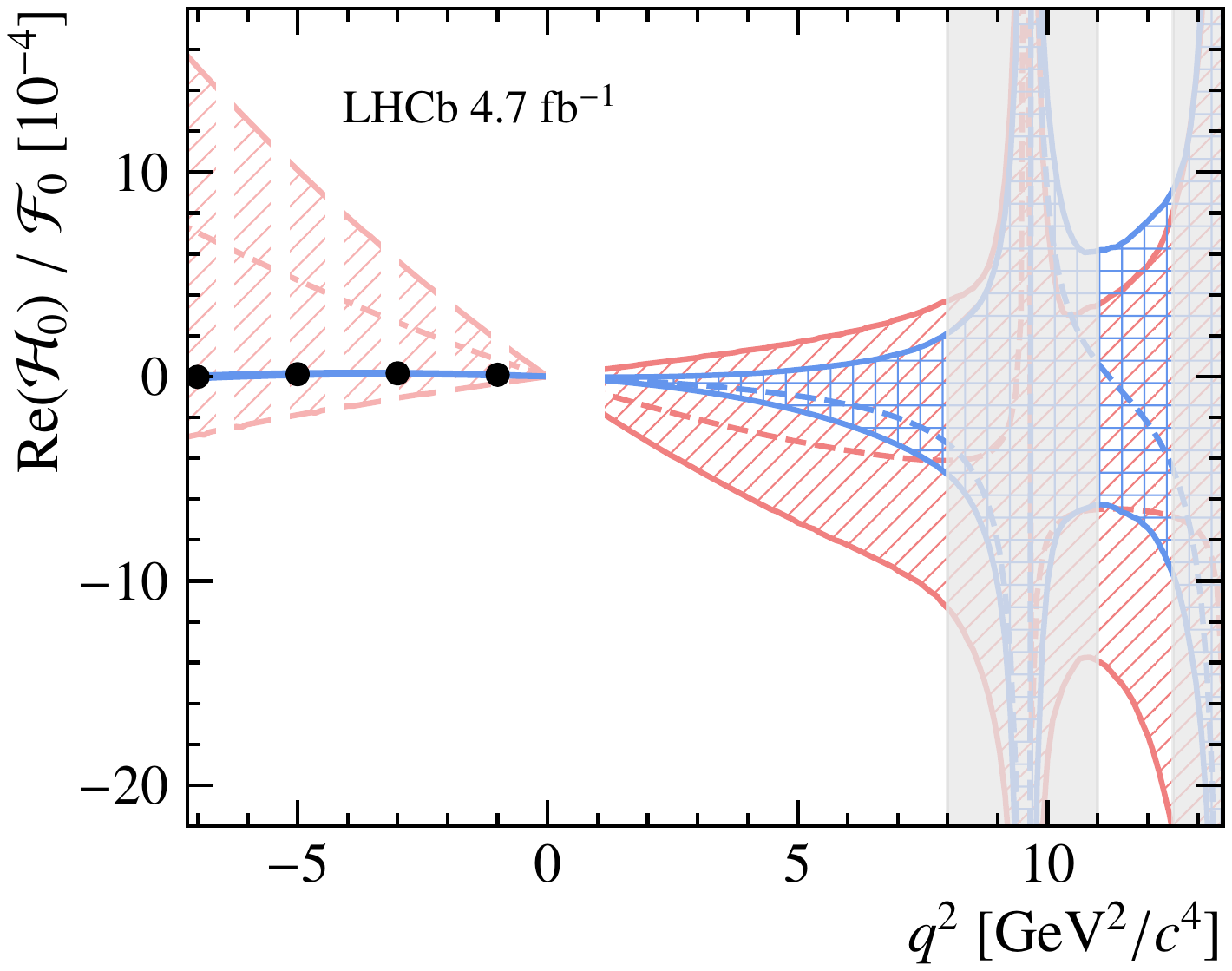}\\
\includegraphics[width = 0.32\textwidth]{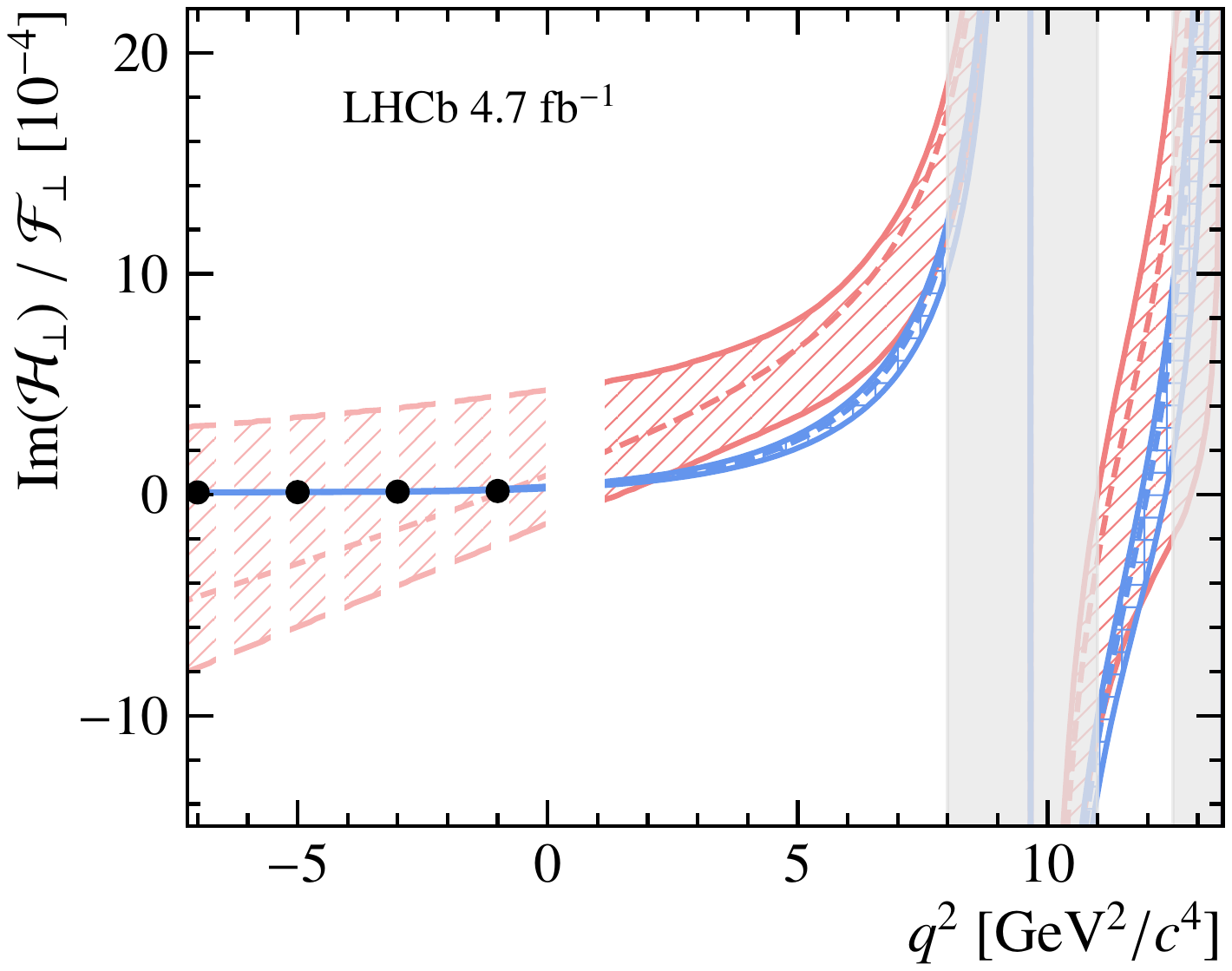}
\includegraphics[width = 0.32\textwidth]{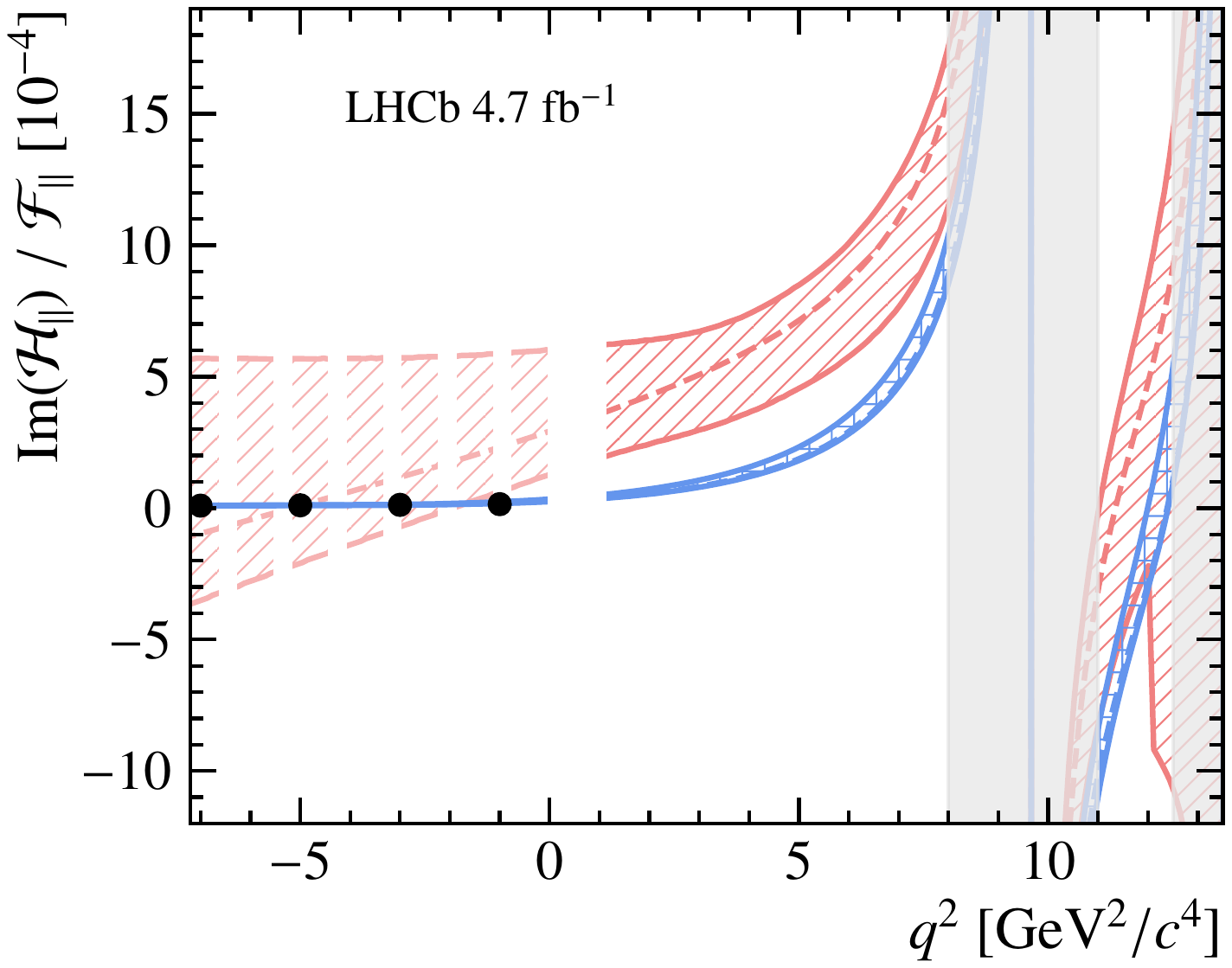}
\includegraphics[width = 0.32\textwidth]{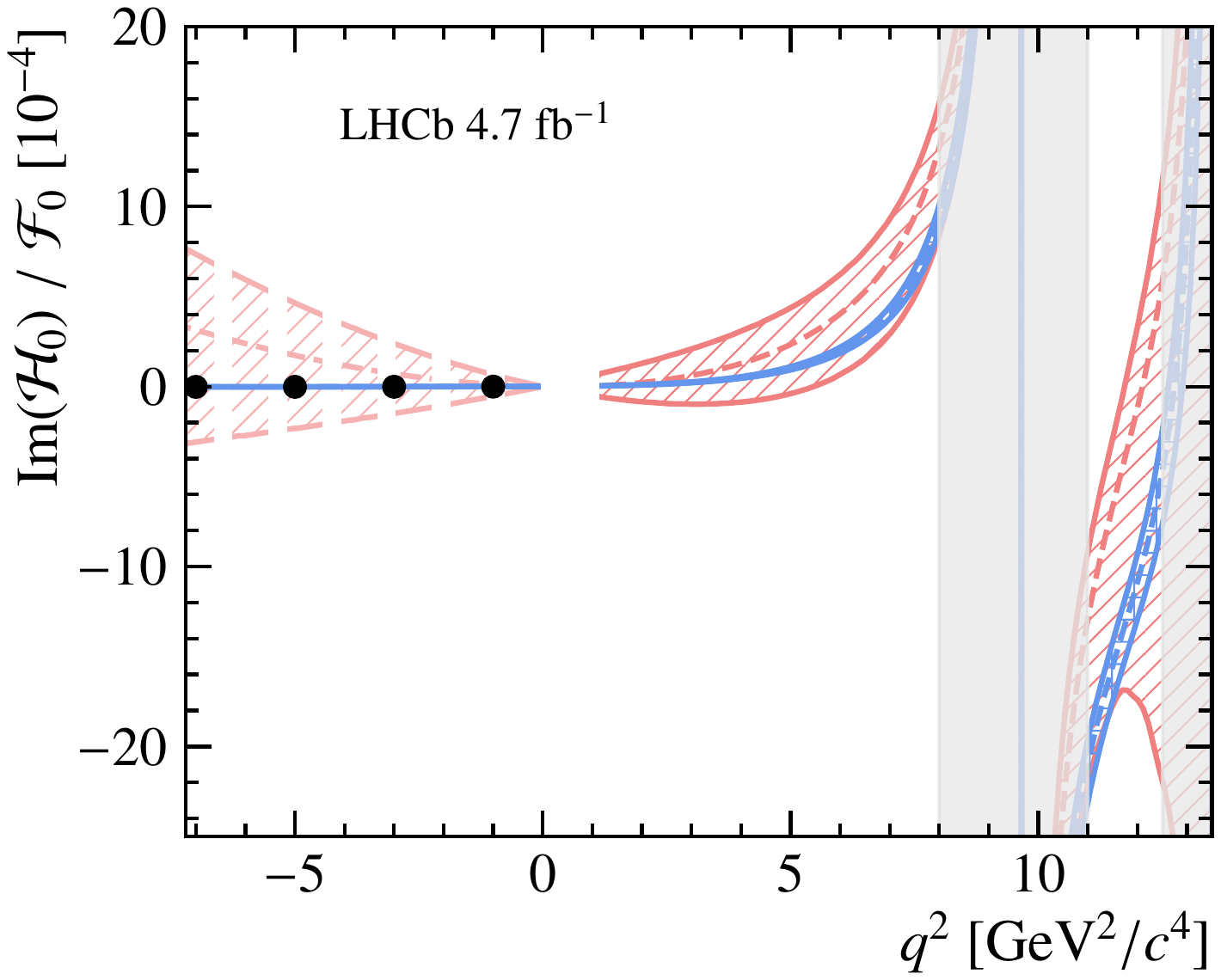}
\caption{The fitted non-local form factors. The blue curves are the fit with the negative \qsq constraints, the red curve is the fit without. The LCSR predictions are shown by the black points. }
\label{sec:results:fig:nonlocalffs}
\end{figure}

The fitted non-local form-factors are shown in Fig.~\ref{sec:results:fig:nonlocalffs}. In general the agreement between the fits with or without the $\qsq<0\,{\rm GeV^{2}}$ predictions is good. The exception is in the imaginary part $\rm{Im}(\mathcal{H}_{\parallel})$ where the predictions provide a very tight constraint, whereas the data prefers to pull up the form-factors.

\begin{figure}[]
\centering
\includegraphics[width = 0.48\textwidth]{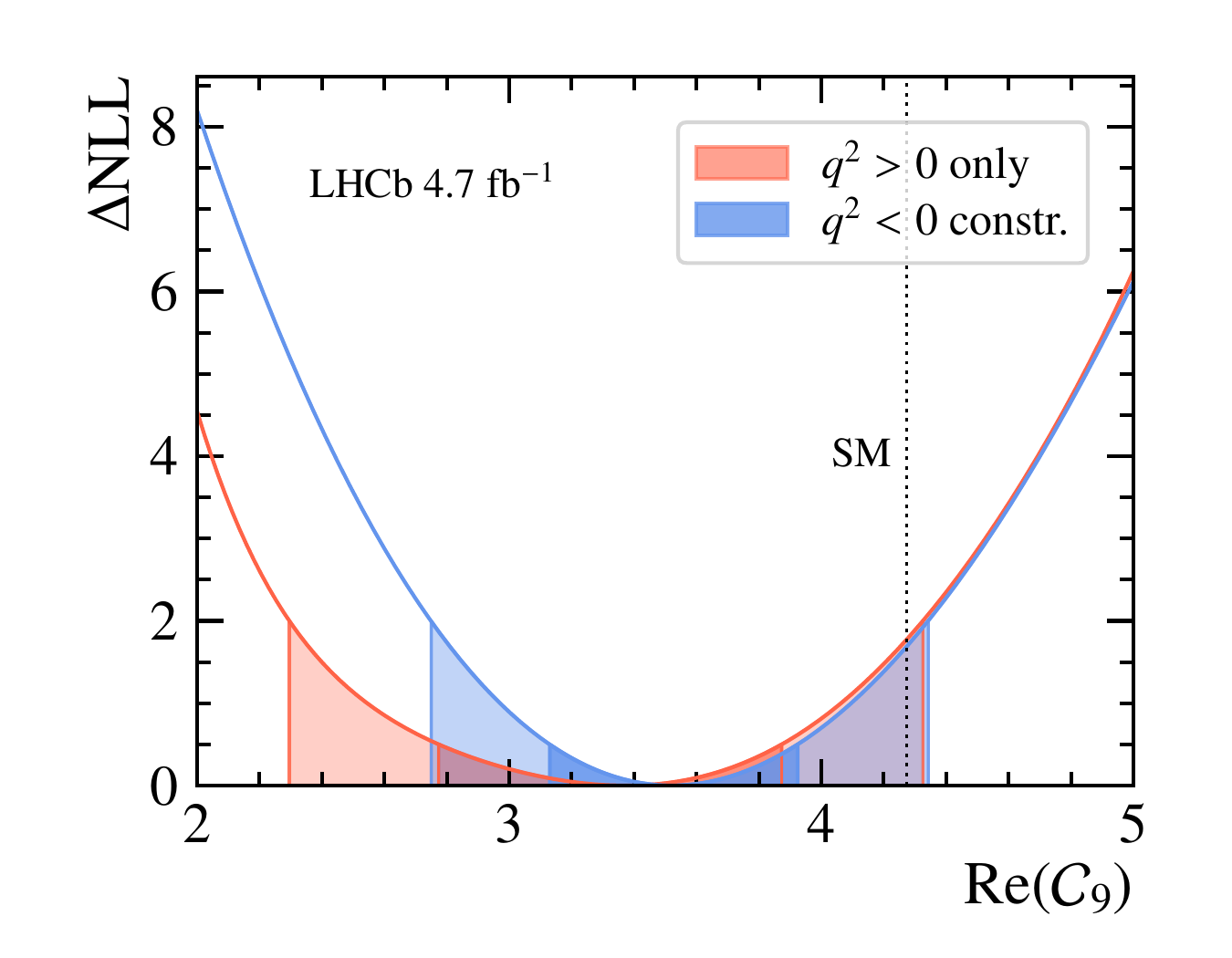}
\includegraphics[width = 0.48\textwidth]{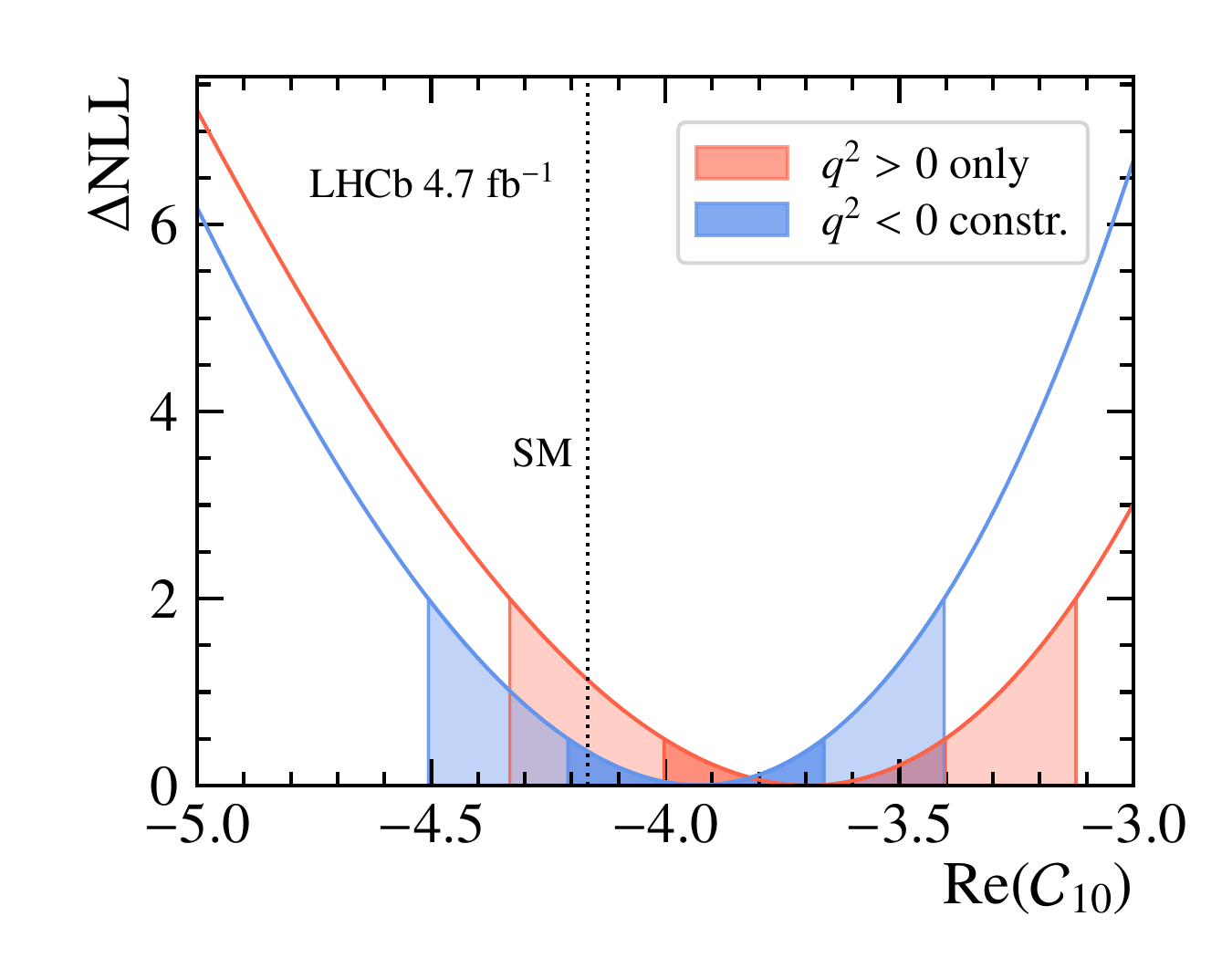}
\caption{One-dimensional profiles of the fitted Wilson coefficients (left) $C_{9}$ and (right) $C_{10}$.}
\label{sec:results:fig:wcs}
\end{figure}

The one dimensional profiles of two of the Wilson coefficients of interest, $C_{9}$ and $C_{10}$, are shown in Fig.~\ref{sec:results:fig:wcs}. The minima of the profiles are consistent with the fit is performed with or without the negative \qsq constraint. The result is that even with a contribution from non local effects the data prefers a value of the Wilson coefficient $C_{9}$ that is away from the SM prediction. When considering all four Wilson coefficients together the best fit point is found to be 1.3(1.4)\,$\sigma$ from the SM when fitting without (with) the negative \qsq predictions.

\subsection{Comparison with binned results}

\begin{figure}[]
\centering
\includegraphics[width = 0.48\textwidth]{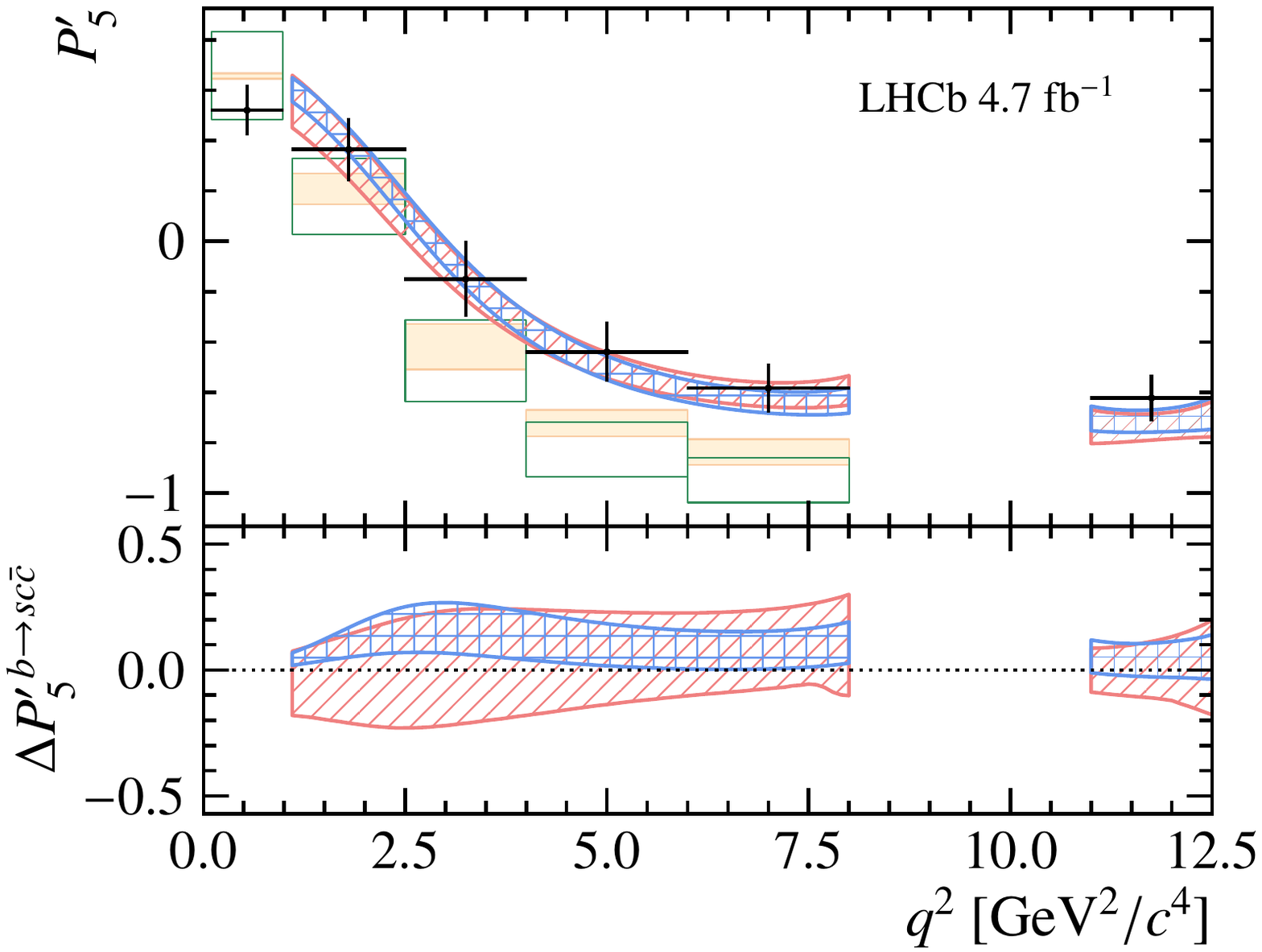}
\includegraphics[width = 0.48\textwidth]{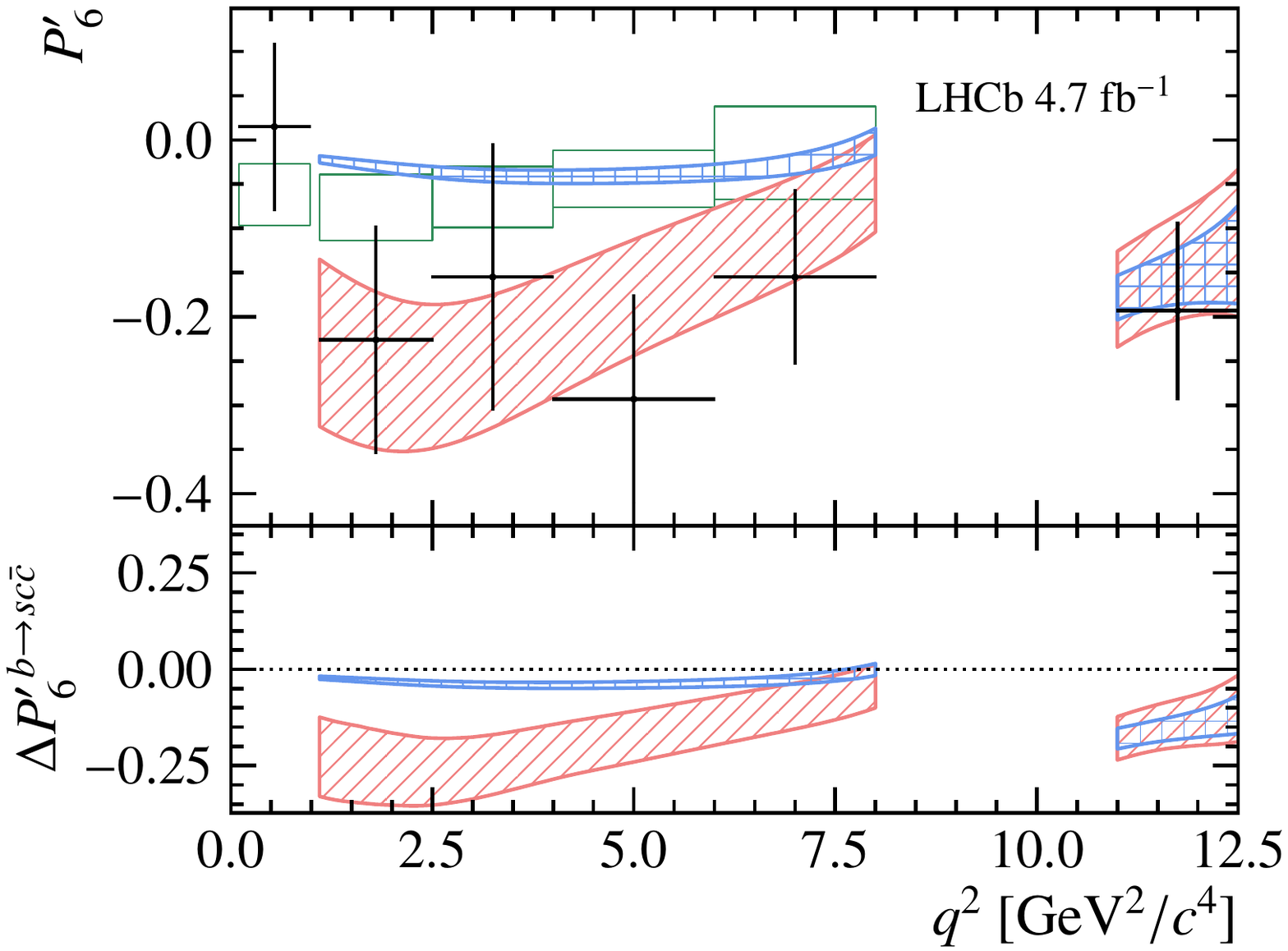}
\caption{Comparison of the results of this analysis with the previously published fit for the binned angular observables~\protect\cite{LHCbbinned}. Shown are (left) $P_{5}^{\prime}$ and (right) $P_{6}^{\prime}$. In the top plots, the black points are the binned results and the blue and red curves are the results of this analysis for the fit (blue) with or (red) without the negative \qsq constraints. The green~\protect\cite{khodjamirian,dmhv} and yellow~\protect\cite{Gubernari_2022} boxes are theory predictions. The bottom plots show the change in the observables when setting the non-local contributions to 0.}
\label{sec:results:comparison:fig:binnedcomp}
\end{figure}

The binned angular observables may be calculated from the fitted amplitudes to provide a direct comparison with the previous LHCb measurement~\cite{LHCbbinned}. In addition, the non-local part of the amplitudes may be set to 0 in this calculation to provide an estimate of the non-local contribution to the measured binned observables. In general the agreement with the published binned analysis is good for the fits with or without the negative \qsq predictions. An example is $P_{5}^{\prime}$, shown in the left of Fig.~\ref{sec:results:comparison:fig:binnedcomp}. The exception is $P_{6}^{\prime}$ (or $S_{7}$ for the non-optimised observables), shown in the right of Fig.~\ref{sec:results:comparison:fig:binnedcomp}. This observable is constructed from the imaginary parts of the amplitudes,
\begin{equation*}
S_{7}\sim \rm{Im}(\mathcal{A}_{0}^{L}\mathcal{A}_{\parallel}^{L\ast} - \mathcal{A}_{0}^{R}\mathcal{A}_{\parallel}^{R\ast}).
\end{equation*}
Consequently the pull of the $\qsq<0\,{\rm GeV^{2}}$ predictions on the imaginary part of the non-local form factors produces a significant shift towards 0 when including them in the fit.

\section{Conclusions}

For the first time an amplitude fit of $\Bz\to\Kstarz\mumu$ has been carried, unbinned in the \qsq variable. The resulting information is complementary to the previously published analyses. The fit has explicitly allowed for a contribution of long-distance $c\bar{c}$ amplitudes to try and disentangle the effects from long-distance and short-distance physics. It is found that even allowing for a long-distance contribution the data prefers the Wilson coefficient $C_{9}$ to be away from the Standard Model.

The LHCb collaboration still has much data in hand that is in the process of being analysed. In particular the binned analysis of $\Bz\to\Kstarz\mumu$ using the full Run~1 and Run~2 data sets is awaited~\cite{binnedpheno}. Furthermore, there is a second unbinned analysis using the full Run~1 and Run~2 data set that has recently been made public. This new analysis also tried to make an estimate of the non-local contributions by using a model of all of the $c\bar{c}$ resonances~\cite{unbinnedone,unbinnedtwo,unbinnedthree}. In addition there is a possibility of making a model-independent determination of the amplitudes of the decay that is continuous in \qsq~\cite{ansatz}.

The LHCb detector has recently started taking data for Run~3. The goal is to collect $7\,\rm{fb}^{-1}$ in 2024 at $\sqrt{s} = 13.6\,\rm{TeV}$ which would represent an equivalent amount of data as that collected in Run~1 and Run~2. A similar harvest is envisaged for 2025 before the next long shutdown of the LHC. This Run~3 dataset will therefore provide a significant increase in precision for the study of these FCNC decay modes.


\section*{References}


\begin{thebibliography}{99}

\bibitem{phimumu}LHCb collaboration, \Journal{\em Phys. Rev. Lett.}{127}{2021}{151801}

\bibitem{lbmumu}LHCb collaboration, \Journal{\em Phys. Rev. Lett.}{131}{2023}{151801}

\bibitem{bfs}LHCb collaboration, \Journal{\em JHEP}{06}{2014}{133}

\bibitem{RKone}LHCb collaboration, \Journal{\em Phys. Rev. Lett.}{131}{2023}{051803}

\bibitem{RKtwo}LHCb collaboration, \Journal{\PRD}{108}{2023}{032002}

\bibitem{LHCbbinned}LHCb collaboration, \Journal{\em Phys. Rev. Lett.}{125}{2020}{011802}

\bibitem{binnedkstplus}LHCb collaboration, \Journal{\em Phys. Rev. Lett.}{126}{2021}{161802}

\bibitem{khodjamirian}A. Khodjamirian {\it et al}, \Journal{\em JHEP}{89}{2010}{89}

\bibitem{dmhv}S. Descotes-Genon {\it et al}, \Journal{\em JHEP}{12}{2014}{125}

\bibitem{eft}G. Buchalla {\it et al}, \Journal{\em Rev. Mod. Phys.}{68}{1996}{1125}

\bibitem{fitone}M. Alguer\'o {\it et al}, \Journal{\em Eur. Phys. J. C}{83}{2023}{648}

\bibitem{fittwo}A. Greljo {\it et al}, \Journal{\em JHEP}{05}{2023}{87}

\bibitem{fitthree}M. Ciuchini {\it et al}, \Journal{\PRD}{107}{2023}{055036}

\bibitem{fitfour}L. Geng {\it et al}, \Journal{\PRD}{104}{2021}{035029}

\bibitem{Gubernari_2022}N. Gubernari {\it et al}, \Journal{\em JHEP}{09}{2022}{133}

\bibitem{ampone}LHCb collaboration, \Journal{\PRD}{109}{2024}{052009}

\bibitem{amptwo}LHCb collaboration, \Journal{\em Phys. Rev. Lett.}{132}{2024}{131801}

\bibitem{Bobeth_2000}C. Bobeth {\it et al}, \Journal{\em Nucl. Phys. B}{574}{2000}{291}

\bibitem{Gorbahn_2005}M. Gorbahn and U. Haisch, \Journal{\em Nucl. Phys. B}{713}{2005}{291}

\bibitem{Horgan_201530}R. Horgan {\it et al}, \Journal{\em PoS LATTICE 2014}{}{2015}{372}

\bibitem{Gubernari_2019}N. Gubernari {\it et al}, \Journal{\em JHEP}{01}{2019}{150}

\bibitem{PhysRevD.90.112009}Belle collaboration, \Journal{\PRD}{90}{2014}{11}

\bibitem{PhysRevD.76.031102}BABAR collaboration, \Journal{\PRD}{76}{2007}{03}

\bibitem{PhysRevD.88.074026}Belle collaboration, \Journal{\PRD}{88}{2013}{07}

\bibitem{PhysRevD.88.052002}LHCb collaboration, \Journal{\PRD}{88}{2013}{05}

\bibitem{Aaij_2012}LHCb collaboration, \Journal{\em Eur. Phys. J. C}{72}{2012}{8}

\bibitem{Gubernari_2021}N. Gubernari {\it et al}, \Journal{\em JHEP}{02}{2021}{88}

\bibitem{binnedpheno}M. Alguer\'o {\it et al}, \Journal{\em JHEP}{12}{2021}{85}

\bibitem{unbinnedone}T. Blake {\it et al}, \Journal{\em Eur. Phys. J. C}{78}{2018}{6}

\bibitem{unbinnedtwo}C. Cornella {\it et al}, \Journal{\em Eur. Phys. J. C}{80}{2020}{12}

\bibitem{unbinnedthree}LHCb collaboration, LHCb-PAPER-2024-011

\bibitem{ansatz}U. Egede {\it et al}, \Journal{\em JHEP}{84}{2015}{6}

\end{thebibliography}
\end{document}